\documentclass[letterpaper]{article}

\usepackage{natbib,alifeconf}  
\usepackage[subrefformat=parens]{subcaption}
\usepackage[ruled]{algorithm2e}

\usepackage{listings}
\title{Simulation of Gacs' Automaton}
\author{Atsushi Masumori*$^{1}$, Lana Sinapayen*$^{1}$ \and Takashi Ikegami$^1$ \\
\mbox{}\\
$^1$The University of Tokyo \\
masumori@sacral.c.u-tokyo.ac.jp\\\\
**These authors contributed equally to this work.
} 

\begin{document}
\maketitle

\begin{abstract}
Peter Gacs proposed a one-dimensional cellular automaton capable of a robust self-reproduction. Because the automaton is exceptionally large and complicated, very few people have ever succeeded in simulating it on a computer or analyzing its behavior. Here we demonstrate a partial simulation of Gacs' automaton (of Gray's version), discussing its robustness. We also discuss the potential applications of Gacs’ framework.

\end{abstract}

\section{Introduction}

Robust self-replication in living systems is still an unsolved problem. Von Neumann developed a theory of self-reproduction by building a self-replicating two-dimensional cellular automaton (\cite{Neumann1966}). This automaton, however, is very unstable against noise that changes the state of the automaton. Cellular automata, such as Conway's Game of Life, are generally vulnerable to noise.

Regarding this stability problem, Peter Gacs showed that there exists a one-dimensional cellular automaton that can be stabilized against any degree of noise (\cite{Gacs2001}). In this automaton, the dynamic evolution of states is discussed as self-regeneration rather than self-replication, which means that the system continues to regenerate itself through self-simulation. A one-dimensional automaton is a system in which cells are linearly arranged, and each cell determines its next state from its current state and the state of its neighbors. For one-dimensional cellular automata with a finite interaction distance (i.e. the number of neighbors), it was considered that even with slight noise, the system becomes unstable and it is not possible to maintain a specific pattern without convergence to one stationary measure. This is called the Positive Rate Conjecture. Gacs falsified the conjecture by building a very special cellular automaton that becomes stable against any amount of noise.

However, since this automaton is large and the rules are complicated, it has been considered difficult to implement and run the automaton on a computer, or analyze it (\cite{DeSa1992}). With the help of recent advances of computer, we implement and analyze Gacs' automaton, we report the progress of the simulation and quantitative results.

\section{Gray's automaton}
\begin{figure}[h]
\centering
\includegraphics[width=7.0cm]{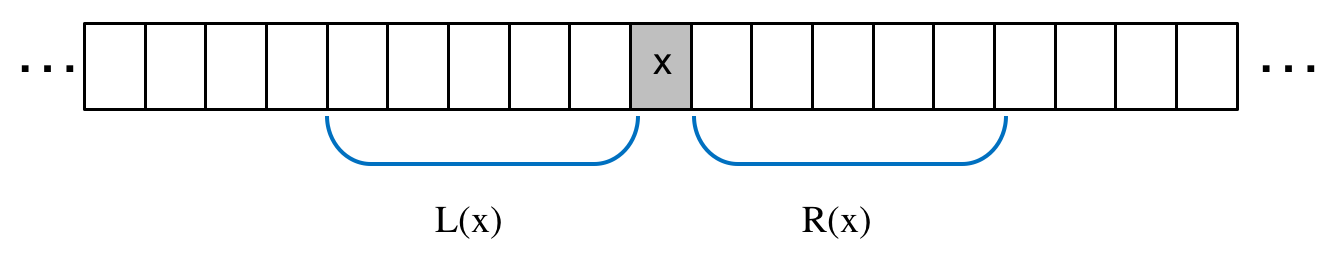}
\caption{Interaction distance in Gray's automaton.}
\label{fig:neighbors}
\end{figure}

We used Gray's automaton, a simple version of Gacs' automaton. Gray's example was introduced as a ``reader's guide'' to Gacs' work in the same journal (\cite{Gray2001}). The automaton consists of colonies which each has $Q$ cells. Although, in Gacs' automaton one cell can only interact with its nearest neighbors, in the Gray's automaton, one cell can interact with five-range neighbors; e.g. the cell at the site $x$ can interact with five left and five right neighbors: $L(x) = \{x-1, x-2, x-3, x-4, x-5\}$, $R(x) = \{x+1, x+2, x+3, x+4, x+5\}$ (Fig. \ref{fig:neighbors}).

\begin{figure}[h]
\centering
\includegraphics[width=7.0cm]{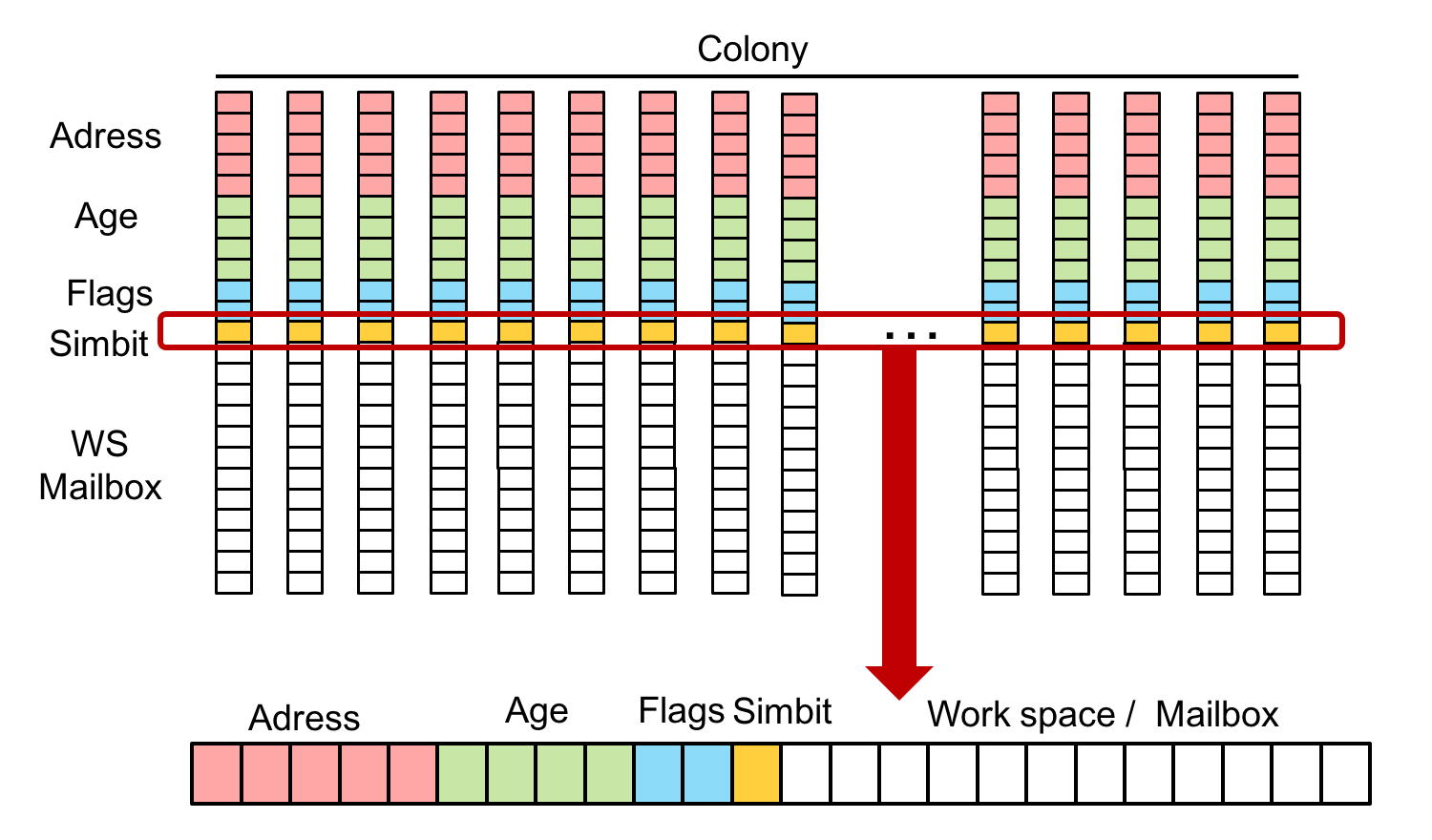}
\caption{Each cell consists of six fields; Age, Address, Flags, SimBit, WorkSpace and MailBox. The set of the first bit of SimBit at all cells in a colony represents the state of a single simulated cell in upper layer.}
\label{fig:concept}
\end{figure}

\begin{figure}[htbp]
\centering
\includegraphics[width=7.0cm]{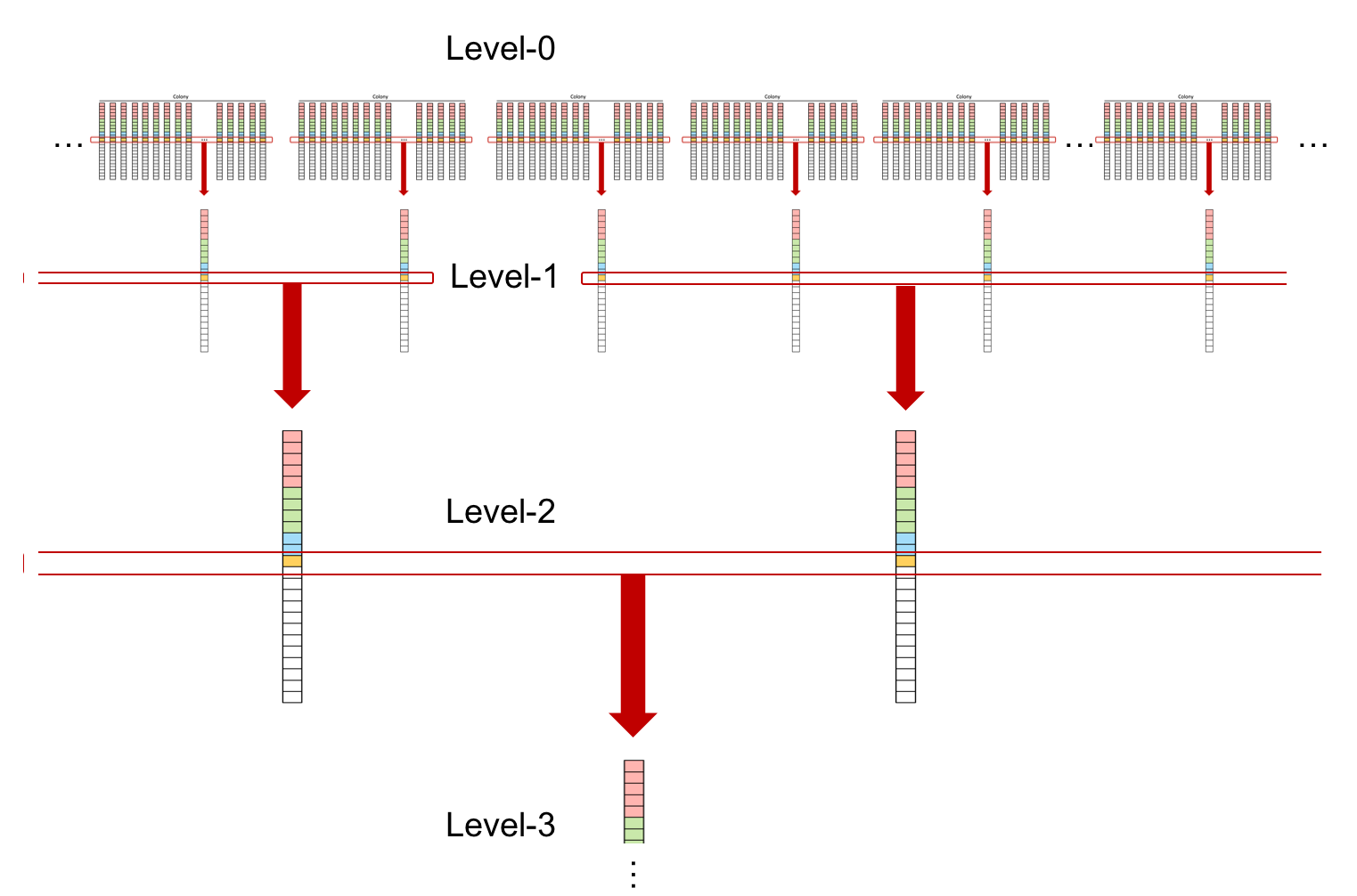}
\caption{The set of the first bit of SimBit in a simulated colony represents a simulated cell in further upper layer.}
\label{fig:self-simulation}
\end{figure}

Each cell consists of six fields; Age, Address, Flags, SimBit, WorkSpace and MailBox. The first two fields are called the local structure, while last three fields are called the simulation structure. The Address field is an integer between 0 and $Q-1$ and the value of the Addess field at the $i$ th cell is $i$ (mod $Q$) in the initial configuration. The automaton should maintain this initial configuration. The Age field counts elapsed time steps, and is reset to 0 when it reaches the value of $U-1$, with $U=128*Q$. The Flag field is composed of Flag1 and Flag2, and both have two states (true or false). The Flags are used to control a process of error correction in the local structure. The SimBit consists of five bits. The set of the first of these bits at all cells in a colony represents the state of a single simulated cell in the upper layer (Fig. \ref{fig:concept}). The simulated cell also has a SimBit, thus the set of the first bit of the SimBit in a simulated colony represents a simulated cell in further upper layer (Fig. \ref{fig:self-simulation}).  The other four bits in SimBit are used for error correction. The WorkSpace stores bits for self-simulation;  the MailBox has a bit for communication with other cells, and all of the bits are set to 0 in the initial configuration. In our simulation, each cell is composed of 293 bits and has $2^{293}$ possible states. The one-dimensional space is divided into $N$ colonies, each composed of $Q$ cell sites.

The summary of the simulation steps is: first, check the consistency of patterns in each colony, and repair the colony if it is broken. If the broken part is too large to be fixed, a coarse-graining process will be useful to correct the large error: each colony corresponds to a single cell state, and these simulated cells will be treated as if they are the normal cell sites and will be corrected. The information in the upper layer trickles down to the lower layer. To correct larger errors, further coarse-graining will be required and information from the simulated layer further above this one will be used for the error correction.

Small errors (level-1 or less) can be corrected by transition rules for only local structure without the self-simulation  (\cite{Gray2001}). In this paper, we report preliminary results without implementing the self-simulation part. Below, we explain the transition rules for error-correction in the local structure except for some transition rules or conditions required for self-simulation.

\begin{figure}[h]
\centering
\includegraphics[width=8.5cm]{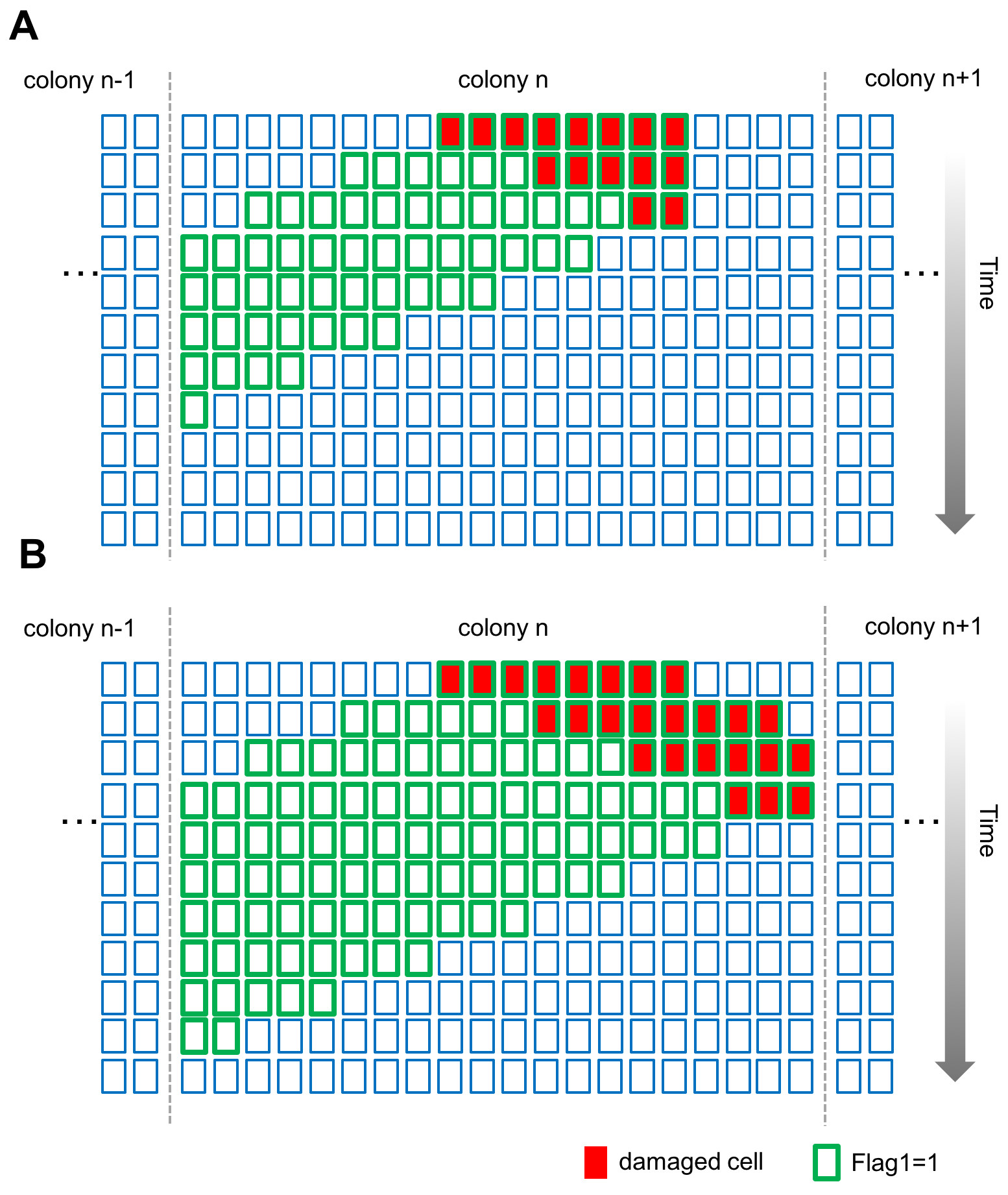}
\caption{Examples of dynamics of error correction in the local structure. A: Errors only happen in the first step. B: Errors ocur even after the first step.}
\label{fig:error_correction}
\end{figure}

The error correction in the local structure is accomplished based on majority vote. If $C(x)$ exists, and Flag1 = 0, the majority vote is executed in $R(x)$. Otherwise, the majority vote is executed in $L(x)$. Here, $C(x)$ represents an apparent colony. If at least three cells of $R(x)$ have values in the Address field that are consistent with each other, an apparent colony $C(x)$ is said to exist. $C(x)$ is defined as the interval of $Q$ cells that includes $x$. If there are no such three cells in $R(x)$, $C(x)$ does not exist (e.g. if the Address values are as $R(x)=\{0, 1, 29, 3, 81\}$, then $C(x)$ exists. If the Address values are as $R(x)=\{3, 4, 6, 5, 2\}$, then $C(x)$ does not exist.) 

The transition rules for Address and Age at a site $x$ are determined by the majority votes at certain cells following the above conditions, and the Age value determined by the vote is incremented by 1 (mod $U$). The original rules for computing the Address, and the rules of transition for the Age have some problems and we modified the rules (see Appendix).

Here, it is defined that there is inconsistency at x, if one of the following four conditions is satisfied: (i) $C(x)$ does not exist. (ii) $C(x)$ exists, and at least three cells in $L(x)\cap C(x)$ have errors in the Address. (iii) at least three cells in $R(x)$ have  different Age values. (iv) at least three cells in $R(x)$ have the same Age value $a$ and at least three cells in $L(x)\cap C(x)$ have different Age values from $a$.

The transition rules for Flag1are defined: the value goes from 0 to1, if one or more of the following three conditions are satisfied: (i) $x$ has an inconsistency. (ii) Flag1 = 1 in at least three cells in $R(x)$. The value goes from 1 to 0, if the following conditions are satisfied: (a) (i) does not hold. (b) no more than one cell in $R(x)\cap C(x)$ has Flag1 = 1. 

The transition rules for Flag2 are defined: the value goes from 0 to 1 if at least one of the following four conditions is satisfied: (i) at least four cells in $L(x)\cap C(x)$ have Flag2 equal to 1, or (ii) the computed value of Flag1 at $x$ is 1 and at least four cells in $L(x)$ have Flag2 equal to 1, or (iii) $C(x)$ does not exist and the computed value of Age at $x$ is divisible by 16.  the value goes from 1 to 0, if condition (iii) is not satisfied and at least one of the following conditions is satisfied: (a) the computed value of Flag1 = 0 at $x$ and there is no cell in $L(x)\cap C(x)$ has Flag2 = 0, or (b) the computed value of Flag1 = 1 at $x$ and no cell in $L(x)$ has Flag2 = 1.

According to these rules, small errors can be corrected like the examples in Fig. \ref{fig:error_correction}. Here, red cells represents damaged cells and green frames represents cells with Flag1 = 1. Fig. \ref{fig:error_correction}A shows the case in which errors only happen in the first step. In such a case, the error island does not expand and shrinks from the left side, and the wave of Flag1 = 1 flows to the left side until it reaches the border between colonies. The Flag1 = 1 island then shrinks from the right. On the other hand, if other errors have occurred during error correction, then sometimes the error expands like in Fig. \ref{fig:error_correction}B, but it does not exceed the border between colonies. If the error is not very large, and is stuck at the right border for long enough, then the error correction process from the left side catches up with the right side. The dynamics of the Flag1 are as in the first example.

\section{Results}
We report preliminary results without implementing the self-simulation part, where $Q$ is set to 271. We demonstrate that the automaton can correct errors with high probability even without self-simulation. In these experiments, at each time step, a number of cells are destroyed according to the error rate. The entire contents of the cell is replaced with a random bit string. . 

\begin{figure}[htbp]
\centering
\includegraphics[width=7.0cm]{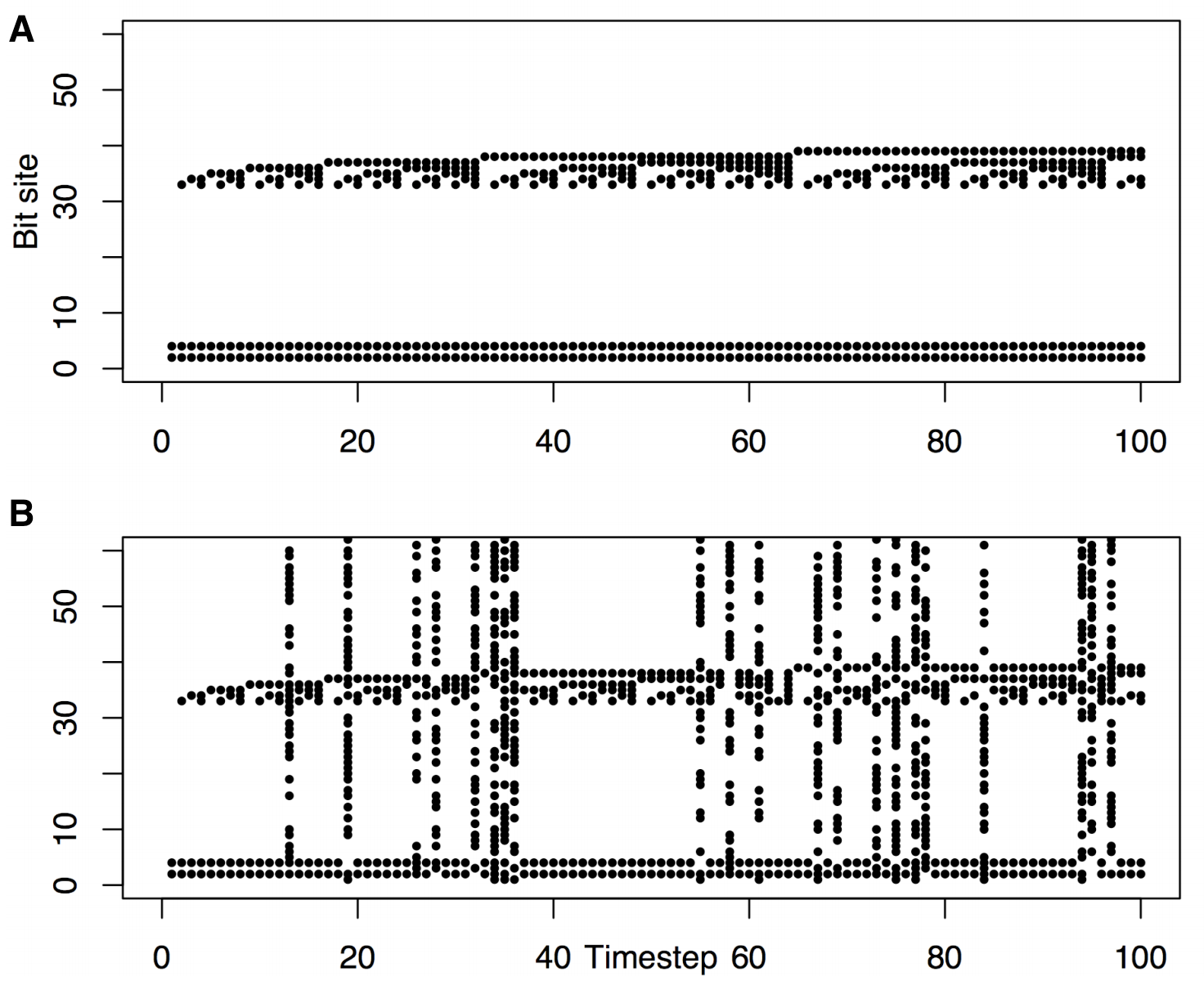}
\caption{The horizontal axis is time, the vertical axis is a state space (Age bit). A: no noise. B: self-correction of bits damaged by noise.}
\label{fig:snapshot}
\end{figure}

\begin{figure}[h]
\centering
\includegraphics[width=7.0cm]{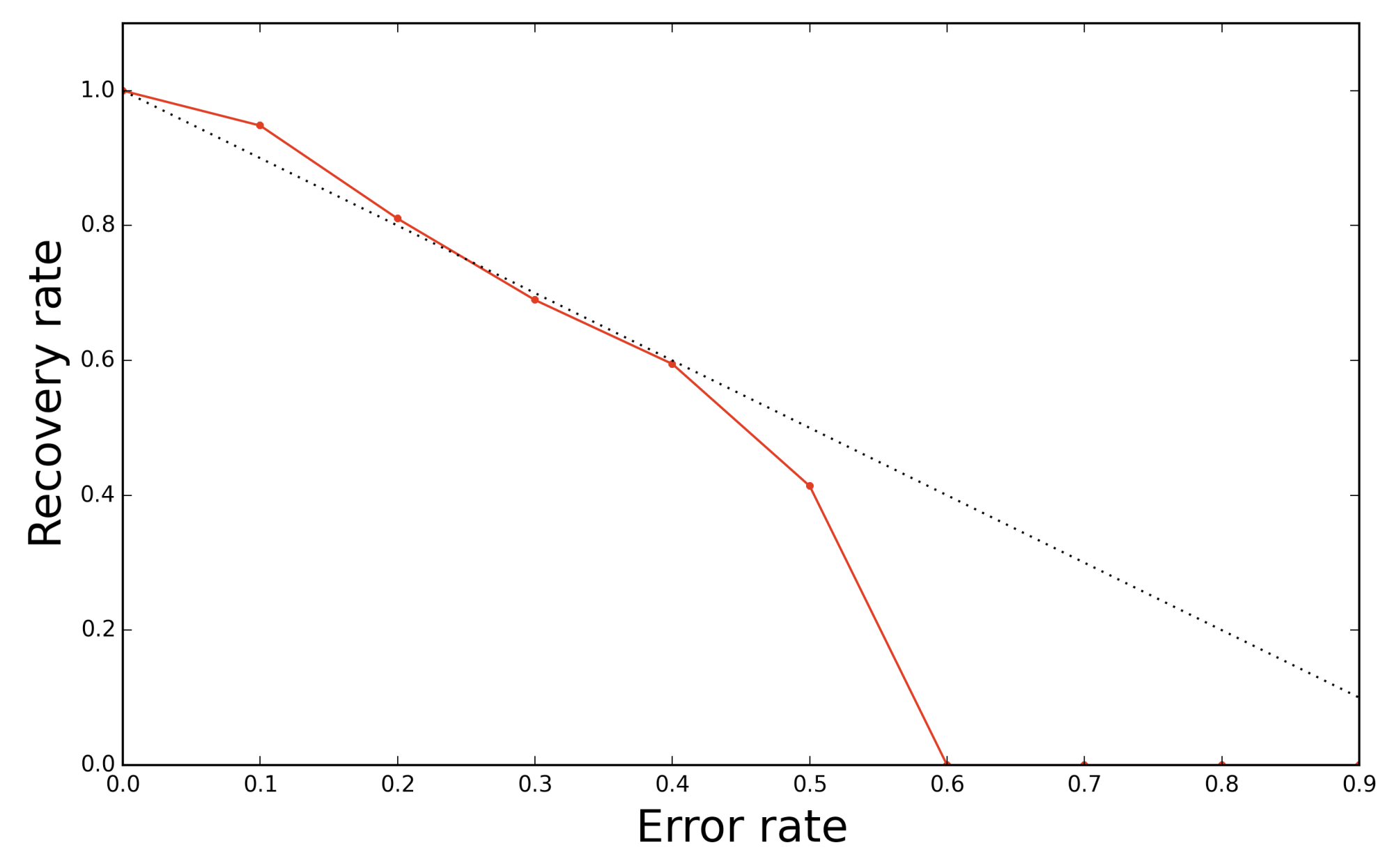}
\caption{Recovery rate against errors. The horizontal axis shows the error rate and the vertical axis shows the recovery rate.}
\label{fig:recoveryrate}
\end{figure}

Fig. \ref{fig:snapshot} is a snapshot of spatio-temporal pattern showing how the errors are getting corrected. The horizontal axis is time, the vertical axis is the state space. Fig. \ref{fig:snapshot}A shows the state without noise and Fig. \ref{fig:snapshot}B shows the behavior of error correction. As you can see from this figure, even if the error rate is high, the automaton can correct them well. Fig. \ref{fig:recoveryrate} shows the recovery rate for different error rates.The recovery rate is high even for high error rates, but at some point (error rate is more than 60\%) the error correction fails and the automaton cannot restore its original state. The simulation result shows that the recovery rate is quite better than we expected, even without self-simulation.

\begin{figure}[h!]
\centering
\includegraphics[width=5.8cm]{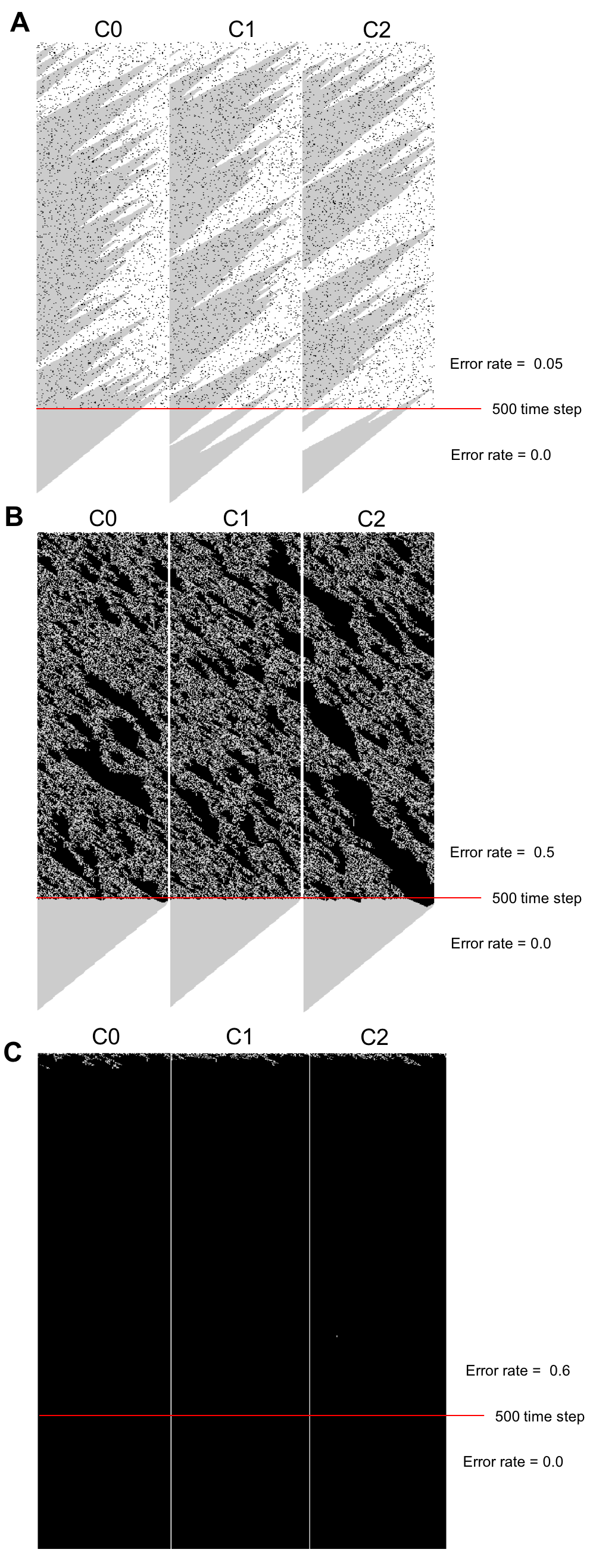}
\caption{Error correcting dynamics. A black dot represents a damaged cell, and a gray dot represents a cell with Flag1 = 1. There is noise with specified probabiliy during 500 time steps and no noise after 500 steps. The dynamics of colonies 0 to 3 are shown. A: Error rate = 0.05. B: Error rate = 0.5. C: Error rate = 0.6.}
\label{fig:pattern}
\end{figure}

We also conducted the other simulation experiments to see the dynamics of error-correction where we input noise to each cell with a certain probability during 500 time steps and remove the noise input after 500 steps. Fig. \ref{fig:pattern} shows the examples of  error correction for each error rate. Here, a black dot represents a damaged cell site, and a gray dot represents a cell with Flag1 = 1, broadly implying that an error was detected in the cell’s state or in the neighboring cells.

As we can see from the simulation with an error rate of 0.05 (Fig. \ref{fig:pattern}A), for small errors, the error correction process is almost always completed in 1 step by the majority vote with the neighbors.The information of Flag1 = 1 propagates to the cell on the left side, and if there is no inconsistency at the border of colony, the wave of Flag1 = 1 stops, and the original state will be completely restored.

As we can see from the case of the error rate of 0.5 (Fig. \ref{fig:pattern}B), the error correction is not completed in 1 step and rather, the island of error temporarily becomes large. However, the islands gradually shrinks and finally fades away while flowing to the right. The island of error goes to the right side as a result of the transition rules of the automaton. 

Even in the case of error rate of 0.5, if the noise input is removed after 500 time steps, the automaton is restored to its original state. On the other hand, when the error rate is 0.6 or more, error correction does not work well, and the original state can not be restored even when noise input is removed after 500 time steps (Fig. \ref{fig:pattern}C).

\begin{figure}[h]
\centering
\includegraphics[width=7.0cm]{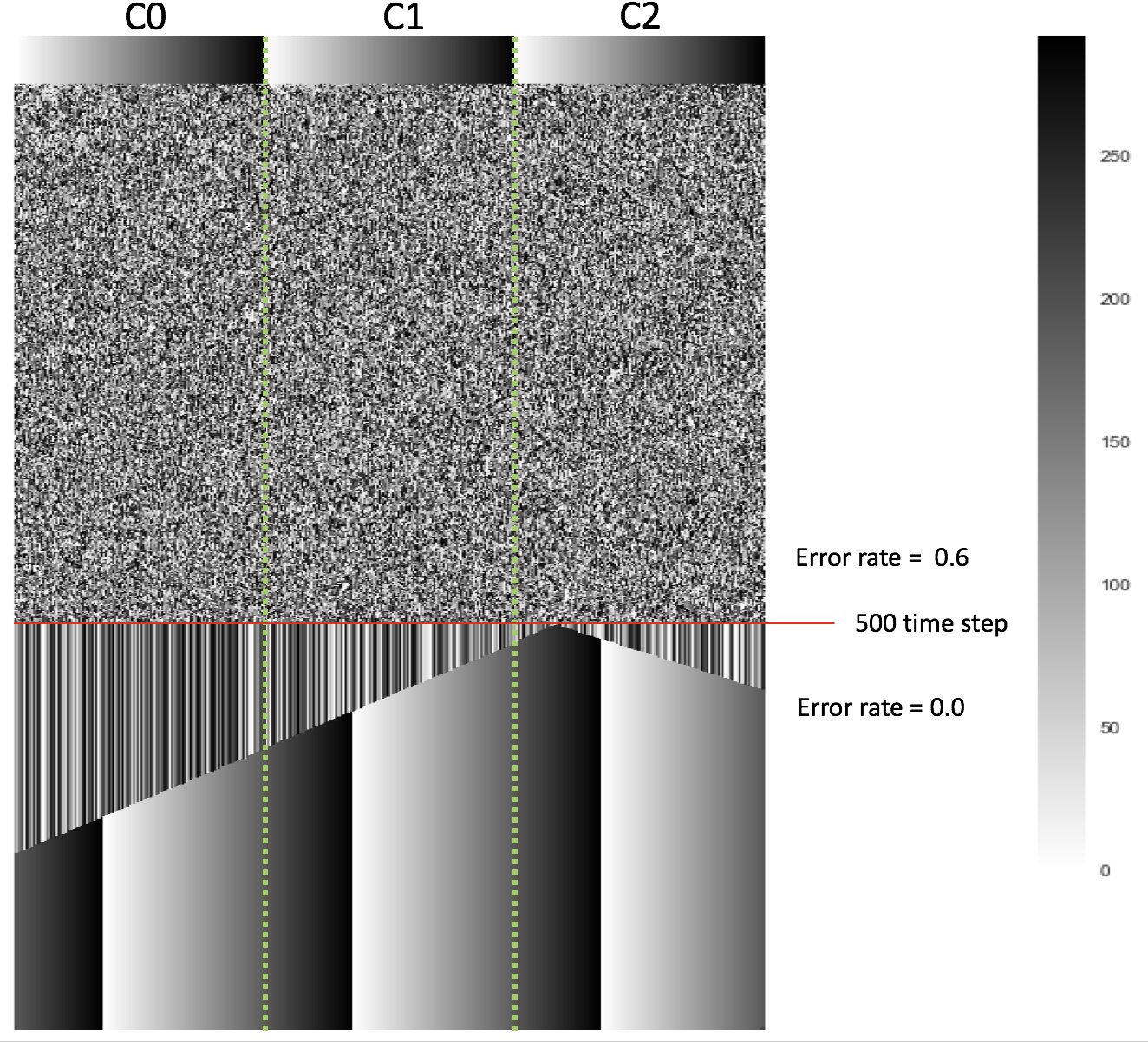}
\caption{Phase shifting observed in the case of error rate = 0.6. The Address values of all cells (0 to 270) are displayed in gray scale. There is noise with error rate = 0.6 during 500 time steps and no noise after 500 steps.}
\label{fig:phaseshift}
\end{figure}

Interestingly, as shown in Fig. \ref{fig:phaseshift} , in some cases with an error rate of 0.6, although different from the original state, the automaton can be restored in a state where the Address of each cell is shifted. Here, the Address values of all cells (0-270) are displayed in gray scale. After 500 time steps, all addresses are in the right order but they are shifted to the right. This seems to be a phase transition.

\section{Discussion}
The concept of robustness is necessary to construct artificial life systems in the real world. Without robustness, cellular automata models are just conceptual ideas that never stand up in the harsh world. The simulation here focuses on a one-dimensional cellular automaton that holds a simple static pattern, but this research can be developed in two directions.

First, it can be developed to focus more on the self-simulation aspect of living systems. From the cell to brain systems, systems can self-reproduce in a closed form. This is called autopoiesis (\cite{Varela1979}). To update the discussion of autopoiesis and discuss a sort of open-autopoiesis idea, we might get insight from the structures of Gacs' large and complex automaton. Gacs' idea can be applicable to brain systems. Self-consciousness in the brain is constantly monitoring and simulating its own state, like Gacs' self-simulation. We consider that Gacs' system can be translated into neural circuits. There are several neural network models that simulate environments to predict and to select actions, but there are few models that simulate themsleves (without falling into infinite regression).

Secondly, Gacs' idea can be connected with robust pattern recognition where we can extract meaning patterns by removing noise. Recently, robust pattern recognition has been developed by deep learning. If we consider robust visual perception as a problem of how to restore a pattern from a noisy pattern, self-simulation can lead to the same results as predictive coding. 

The theory of robust perception can also lead to a model of perception in living systems.
Data massively flowing to complex systems can lead to make the system life-like \cite{Ikegami2013}. We believe Gacs' automaton will also contribute to this research theme.

\section{Acknowledgements}
This work was partially supported by MEXT as Post-K Computer Exploratory Challenges---Construction of models for interaction among multiple socioeconomic phenomena (hp170272).

\appendix
\section{Appendix: Problems in Gray's automaton.}
We found some errors in the Reader's guide (\cite{Gray2001}) and corrected or added rules as follows.

\subsection{Problem in computing Address value}
Grays explains how to use a cell's neighbors to find what that cell's address should be and check this ideal address value against the address value that the cell currently holds.

\subsubsection{Original rule}
p.22: ``if the Address field at some site y is involved in a majority vote that is being used to determine the new Address value at a site x, then the Address value at y needs to be adjusted by adding $(x-y)~mod~Q$'' can be summarized as:
$adjusted\_address\_y = (address\_y + x - y)~mod~Q$.

\subsubsection{Issue}
This requires to know $x$ and $y$ (the position of the cell and its neighbor) with certainty, but if we knew these values we wouldn't need to calculate addresses in the first place. We could just say $address\_x = x$. We do not know $x$ and $y$ so this rule cannot be used. 

\subsubsection{Modified rule}~\\~
$adjusted\_address\_y = (address\_y - i\_y)~mod~Q$,~\\
where $i$ is the index of the neighbor relative to this cell ($i$ in $\{-5..-1\}\cup\{1..5\}$). The only data we know with certainty is which neighbor we are accessing ($i\_y$), so we use this information.

For example, in a case where there are no errors,
if\\ $address\_x = 8$, $address\_y = 1$, $Q = 10$\\ (colonies aligned as $[0,1,2,3,4,5,6,7,x,9][0,y,z,...]$), then\\ $i\_y = 3$ ($y$ is the 3rd neighbor to the right of $x$),\\ $adjusted\_address\_y = (1-3)~mod~10$ $= -2~mod~10 = 8$.\\
We can do the same with $z$:\\ $i\_z = 4$ \\ $adjusted\_address\_z = (2-4)~mod~10 = -2 mod 10 = 8$.

The adjusted addresses equal the correct address for $x$, $address\_x$.
If $x$ held an erroneous address, for example $address\_x = 6$ instead of the expected value 8,\\ 
$adjusted\_address\_y = (1-3)~mod~10 = -2~mod~Q = 8$\\
$adjusted\_address\_z = (2-4)~mod~10 = -2~mod~Q = 8$.

The rule is still working. The final corrected $address\_x$ is obtained by taking the majority in the 5 neighbor's $adjusted\_address$ values, so that even if some neighbors hold wrong addresses, $x$ can recover.

\subsubsection{Sample code (Java)} : 
\begin{lstlisting}[language=Java]
//calculate adjusted addresses
int adj_addr[] = new int[5];
for(int i=0; i<s;i++){
   //the adjusted address
   adj_addr[i] = neighbors[i].getAddr() - (i + 1)*direction;
   
   // java does not have a modulo 
   // operator that works with negative 
   // numbers.
   if(adj_addr[i]<0) {
   	adj_addr[i]+=Constants.Q;
   }
   if(adj_addr[i]>=Constants.Q){
   	adj_addr[i]-=Constants.Q;
   }
}
\end{lstlisting}


\subsection{Problem in transition rules for Flag2 and Age}
\subsubsection{Original rule}
p.21: Partial transition rule for Flag2: 
``0$\to$1 if at least one of the following four conditions is satisfied: (i) at least four sites in $L(x)\cap C(x)$ have Flag2 equal to 1, or (ii) the computed value of Flag1 at $x$ is 1 and at least four sites in $L(x)$ have Flag2 equal to 1, or (iii) $C(x)$ does not exist and the computed value of Age at $x$ is divisible by 16, or (iv) at least three sites in $N(x)\cap C(x)$ have Workspace.Flag2 equal to 1.''

p.22: Transition rule for Age: 
``If $C(x)$ exists and either the computed value of Flag1 at $x$ is 0 or the computed value of Flag2 at $x$ is 1, take the majority vote in $R(x)$.''

\subsubsection{Issue}
There are circular conditions to compute the Age and Flag2 fields. Condition (iii) for computing Flag2 depends on a computed value of Age, and the computed value of Age depends on the computed value of Flag2.

\subsubsection{Modified rule}
We choose to prioritize the age calculation, and change the Flag2 transition rule to use it with an uncorrected value of age as:\\
if $\lnot C(x)$ and $age~mod~16==0$\\
then $conditionFlag2iii = true$.

        


\subsubsection{Sample code (Java)} : 
	\begin{lstlisting}[language=Java]
//If C(x) exists and either the computed 
//value of Flag1 at x is 0 or the computed 
//value of Flag2 at x is 1,
//take the majority vote in R(x)
/** computed [age, address] */
int[] ageAndAddr;
if(hasApparentColony & 
(!nextState.flag1() | nextState.flag2())){
  ageAndAddr = majorities(r_neigh);
  nextState.setAge(ageAndAddr[0]);
  nextState.setAddr(ageAndAddr[1]);
}else{
  ageAndAddr = majorities(l_neigh);
  nextState.setAge(ageAndAddr[0]);
  nextState.setAddr(ageAndAddr[1]);
}                

//increment age
nextState.ageThisCell();

//Replaced computed value of age
//by actual value of age to 
//avoid circular conditions.
if(!hasApparentColony &
(this.getAge()%16)==0){
  cond_flag2iii = true;
}

\end{lstlisting}

\subsection{Problem in transition rules for Workspace.Flag1 and Workspace.Flag2}

\subsubsection{Original rule}
p.41: Transition rule for Workspace.Flag1 (not used in this paper, but necessary for self-simulation)
``(iii) the computed value of SimBit at the site $(x-A(x))+(Q-3) $ is 1...''

\subsubsection{Issue}
The computed value of SimBit at that site is definitely unaccessible information.
Example:\\
if $Q = 15$, $[0,1,x,3,4,5,6,7,8,9,10,11,12,13,14]$ \\and $computed\_address\_ x= address\_x=2$,\\ 
then site $(x-A(x))+(Q-3) = (2 - 2) + (15-3) = 12$.
Information in the 12th cell is not accessible to cell $x$, which can only read from its five closest neighbors.
The same issue as Workspace.Flag1 occurs for Workspace.Flag2.

\subsubsection{Modified Rule:}
There does not seem to be an easy solution to this issue in Gray's version. Gacs' work does not suffer from this issue, as the Workspace computation only uses information from its own cell and 2 neighbors.

\footnotesize
\bibliographystyle{apalike}
\bibliography{alife2018_gacs} 

\end{document}